\pdfoutput=1
\documentclass[aps,pre, twocolumn, groupedaddress]{revtex4-1}

\usepackage{lipsum}
\usepackage{mathtools}
\usepackage{graphicx}
\usepackage{dcolumn}
\usepackage{amsmath}    
\usepackage{amssymb}
\usepackage{bm}
\usepackage{hyperref}
\usepackage{latexsym}
\usepackage{verbatim}
\usepackage[normalem]{ulem}
\usepackage{setspace}

\usepackage{color}
\usepackage[caption=false]{subfig}
\fontsize{20pt}{20pt}
\def\beq{\begin{equation}}
\def\eeq{\end{equation}}

\def\beq{\begin{equation}}                          
\def\eeq{\end{equation}}                          
\def\bea{\begin{eqnarray}}                          
\def\eea{\end{eqnarray}}

\DeclareRobustCommand{\uvec}[1]{{%
  \ifcsname uvec#1\endcsname
     \csname uvec#1\endcsname
   \else
    \bm{\hat{\mathbf{#1}}}%
   \fi
}}
\preprint{}
\begin{document}
\preprint{}
\title{Domain growth kinetics of active model B with thermal fluctuations}
\author{Shambhavi Dikshit$^{1}$}
\email{shambhavidikshit.rs.phy18@itbhu.ac.in}
\author{Sudipta Pattanayak$^{2}$}
\email{sudipta.pattanayak@cyu.fr}
\author{Shradha Mishra$^{1}$}
\email{smishra.phy@iitbhu.ac.in}
\author{Sanjay Puri$^{3}$}
\email{purijnu@gmail.com}
\affiliation{$^{1}$ Department of Physics, Indian Institute of Technology (BHU) Varanasi, India 221005}
\affiliation{$^{2}$ Laboratoire de Physique Th$\acute{e}$orique et Mod$\acute{e}$lisation, UMR 8089, CY Cergy Paris Universit$\acute{e}$,Cergy-Pontoise, France 95302}
\affiliation{$^{3}$ School of Physical Sciences, Jawaharlal Nehru University, New Delhi–India}

\date{\today}

\begin{abstract}
We perform  a comprehensive study on the role of thermal noise on the ordering kinetics of a collection of active Brownian particles modeled using coarse-grained conserved active model B (AMB). The ordering kinetics of the system is studied for the critical mixture when quenched from high to a low temperature. The structure of the growing domains changes from isolated droplet type for AMB without noise  to bi-continuous type for active model B with noise (AMBN). Unlike the passive counterpart of the AMB, the noise is  relevant for the growth kinetics of  the  AMB. We use extensive numerical study, as well as dynamic scaling hypothesis to characterize the kinetics of the system.  We find that  the asymptotic growth law for AMBN is diffusive Lifshitz-Slyozov (LS) type, whereas  it was reported previously that the asymptotic growth law for the AMB without noise is slower, with a growth exponent $4$. Moreover, the kinetics of the growing domains show a strong time dependent growth for AMBN. 
The growth law shows a crossover from early time $1/3$ value to intermediate time $1/4$ value, and it again traverses  from $1/4$ to $1/3$ asymptotically. The two different scaling functions are found for intermediate time and late time with growth law $1/4$ and  $1/3$ respectively. 

\end{abstract}
\maketitle
\section{Introduction \label{Introduction}}
Recent research has shown interest in addressing the active materials, a system of self-propelled particles that violates the time-reversal symmetry by taking energy from the environment and converting it into its persistent motion at each constituent level \cite{Marchetti2013,Ramaswamy2010, mc12, jyl17}.
Such systems exist in a range from small microscopic scale \cite{Rappel1998, Berg2004} to large macroscopic scale up to a few meters \cite{Vicsek1995, Couzin2005, Ballerini2008}. There have been many theoretical, and experimental studies, that address natural microorganisms such as bacteria \cite{Marchetti2013,mc12} or self-propelled micro-swimmers in various mediums \cite{Howse2007,Ebbens2010,Thutupalli2011,Volpe2011,Palacci2013,Pattanayak2019}. Active Brownian particle (ABP), an example of self-propelled particle, having shape symmetry and preferred direction of motion, constitutes a theoretically idealized model for such systems \cite{Fily2012}. \\
In recent studies, it has been found that ABP shows a directed transport and there is nontrivial dependence of transport speed on the local particle density \cite{Pattanayak2019,Koumakis2013}. 
When such ABPs are placed in collection, the system undergoes a phase separation,  even at low packing fractions. This phase separation is called as
motility induced phase separation (MIPS) and has been studied in various simulations and experiments \cite{Fily2012,Stenhammar2013,Cates2010, Thompson2011,Wysocki2014,pub18,Volpe2011}. \\
The phase separation kinetics in corresponding equilibrium or passive colloidal systems is well studied in theory and is described by dynamical $\phi^4$ - field theories, where $\phi$ is a conserved scalar order parameter field.  The model is named as model B (PMB) \cite{Hohenberg1977}. The theory uses the simplest form of the Cahn-Hiliard (CH) equation and it shows the time dependence of domain growth as power law $L(t) \sim t^{1/3}$ \cite{Cahn1959,Hohenberg1977}.  The analogous model to understand the phase separation in ABP system is called as active model B (AMB). Recent studies are done on AMB by introducing an additional activity dependent term $\lambda$ in the order parameter update equation of Cahn-Hilliard (CH) equation \cite{wts14,sudiptasoft,sudiptapre}, which can not be derived from an equilibrium free energy functional. It is found that the activity slows down the growth kinetics in the active model \cite{sudiptapre}, and the asymptotic growth law shows a crossover from early time LS growth to late time  $1/4$ growth for the athermal systems. However, in the above-mentioned studies, the ABP system is considered as an athermal system, here, we want to emphasize that  in any real system thermal fluctuations play an important role. \\
The ordering kinetics in corresponding equilibrium systems, PMB, remains unaffected in the presence of thermal noise \cite{Bray1994, Bray1990, Brayprl, Puri1988}. However, the density dependence motility in active systems leads  the thermal noise multiplicative in nature \cite{wts14,Stenhammar2013}. Such multiplicative noise can be relevant for the phase separation kinetics in the ABP system due to the dynamic nature clusters. Also, at the boundary of a cluster of ABPs there is always exchange of particles with the bulk, and this leads to the presence of non-zero current at the order-disorder interface, whereas the current in the corresponding passive model is zero at the interface.\\  

In this work, we study the effect of thermal fluctuations on the domain growth kinetics and domain morphology in AMB. We consider the system a critical mixture.  Using extensive numerical study as well as the dynamic scaling hypothesis, we show that the thermal noise is relevant for the Active model B, whereas for the passive model B, it is irrelevant \cite{Bray1994}. We find that the asymptotic growth law for AMBN is diffusive  Lifshitz-Slyozov (LS) type, whereas, for an athermal AMB system, the asymptotic growth is slower, with a growth law of $1/4$ as reported in the previous study  \cite{sudiptapre}.
The kinetics of the growing domains shows a strong time dependent growth law in the presence of thermal noise. The growth law shows a crossover from an early time of $1/3$  to $1/4$ at intermediate times, and it again approaches to $1/3$ at late times. We note two different scaling behavior for the correlation functions corresponding to the intermediate and late times.\\
The paper is organized as follows. In Sec. \ref{model}, we introduce the AMB with multiplicative noise and discuss the numerical and parameter details. In Sec. \ref{Results}, we present  detailed numerical and analytical results for domain growth in the AMB. In Sec. \ref{Discussion}, we summarize our main conclusions.

\section{Model \label{model}}

We consider an assembly of ABPs on a two-dimensional ($d=2$) substrate using the coarse-grained dynamical equation for local order parameter $\psi({\bf{r}},t)$ of particles. The dynamical equation of motion for local density $\psi$ can be obtained as below. The derivation of the hydrodynamic equation can be found in \cite{ct15}. The resultant model can be expressed as a continuity equation for the conserved order parameter:
\begin{equation}
\begin{aligned}
\begin{split}
\frac{\partial}{\partial t} \psi({\bf{r}},t) &= -{\bf{\nabla}} \cdot {\bf{J}}({\bf{r}},t)  +  {\bf{\nabla}} \cdot {\bf n}({\bf{r}},t),   \\
\label{eq:1}
\end{split}
\end{aligned}
\end{equation}
The first term on the right hand side of equation \ref{eq:1} is the current due to the change in local order parameter $\psi$ and second term is  present due to the thermal noise. The current ${\bf{J}}({\bf{r}},t)$ can be obtained by
\begin{equation}
{\bf{J}}({\bf{r}},t) = - {\bf{\nabla}} \mu ({\bf{r}},t)
\label{eq:2}
\end{equation}
For simplicity we take constant mobility.
The chemical potential 
\begin{equation}
\mu({\bf r}, t) =  -\psi + \psi^{3} - \nabla^{2}\psi + \lambda |{\bf{\nabla}} \psi|^{2} 
\label{eqn_activemodelB}
\end{equation}
The chemical potential $\mu$  is the sum of passive ($\mu_{p}$) and active  ($\mu_{a}$) contributions. The passive part is the same as for Model B, $\mu_{p} = -\psi({\bf r},t) + \psi({\bf {r}},t)^{3} - \nabla^2 \psi({\bf r},t)$, and is derived from the free-energy functional $\mathcal{F}(\psi({\bf r}, t))$ of a symmetric Ginzburg-Landau (GL) $\psi^4$-field-theory \cite{Puri2009}. 
The active contribution of the chemical potential $\mu_{a}$  breaks time reversal symmetry  and has strength $\lambda$, which is a relevant parameter in our study. This term cannot be obtained by the derivative of a free energy, and its origin is due to the self-trapping mechanism of particles at finite motility \cite{Cates2013}. This term provides positive feedback that slows down particles in the high-density region, resulting in an accumulation of the particles. The activity strength $\lambda$ depends on the functional form of the density-dependent speed of the ABPs \cite{wts14}.

\begin{figure}[ht]
\centering
    \includegraphics[scale=0.34]{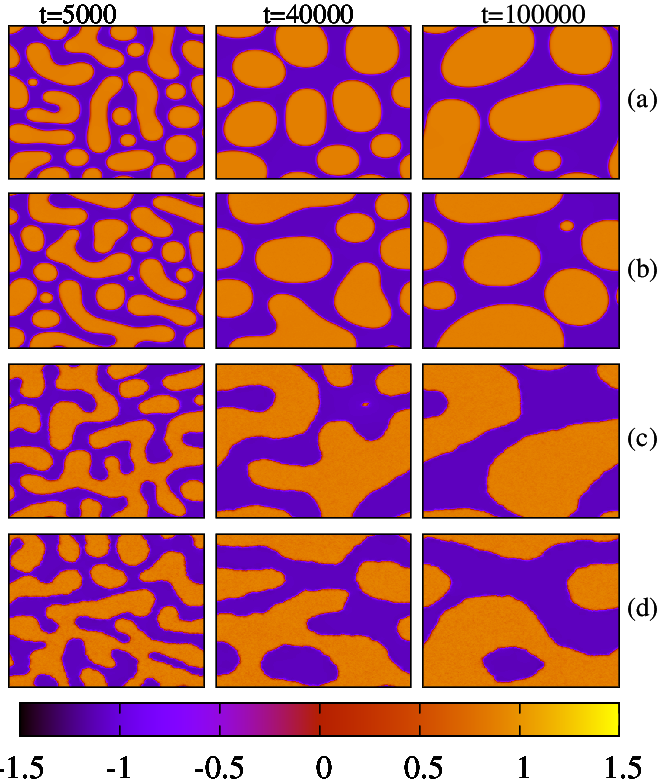}

	\caption{(color online) Evolution snapshots of the system at different times $t$, for various values of the noise strengths $\eta_{a}$. (a-d) are for noise strengths $0$, $0.5$, $0.8$ and $1.0$, respectively. $\lambda=1,$ $L^2=512^2$ and the average order parameter value is $\psi_{0}=0$. The color bar denotes the range of $\psi$ value.}
\label{fig:fig1}
\end{figure}

The above equations are formulated in dimensionless units. These are obtained by rescaling space and time  by a persistence length $v_{0} \tau$ and $\tau$, where $\tau$ is the relaxation time in the system, which we set to $1$. 
 The mean of the thermal noise ${\bf n}({\bf r}, t)$  in Eq. \ref{eqn_activemodelB}  is zero and its two-point correlation is given by, 

\begin{equation}
<{n_{i}({\bf{r}}^{\prime},t^{\prime})n_{j}({\bf{r}}^{\prime\prime},t^{\prime\prime})}>= \eta \delta_{ij}\delta({\bf{r}}^{\prime}-{\bf{r}}^{\prime\prime}) \delta(t^{\prime}-t^{\prime\prime})
\label{eq:4}
\end{equation}
where $<...>$ denotes the average over the probability distribution of the noise ${\bf n}$. 

The strength of the noise $\eta$  in active model is in general multiplicative in nature. 
 The multiplicative nature of noise can be obtained by the local density dependence of motility or by the effective diffusivity $\mathcal{D}(\rho)$  of the particles \cite{Cates2013, Dean1996}. Hence, the form of noise for active model B is  $\sqrt{\mathcal{D}(\rho)\rho}$.  
where, $\rho$ is the local density of active Brownian particles and $\mathcal{D}(\rho) = (1-\rho)^2$ is the effective diffusivity of particles. Furthermore, we can transform the form of noise from local density dependence $\rho$ to local order parameter $\psi$, by $\psi = 2 \rho -1$. 
We also  introduce a parameter $\eta_a$ for the noise such that 
\begin{equation}
\eta = \eta_a \left[(1+\psi({\bf{r}},t))((1-\psi({\bf{r}},t))/2)^2\right]^{1/2}
\label{eq:noise}
    \end{equation}
for $\eta_a = 0$, the effect of thermal noise is turned off.\\
\begin{figure}[ht]
\centering
  \includegraphics[scale=0.20]{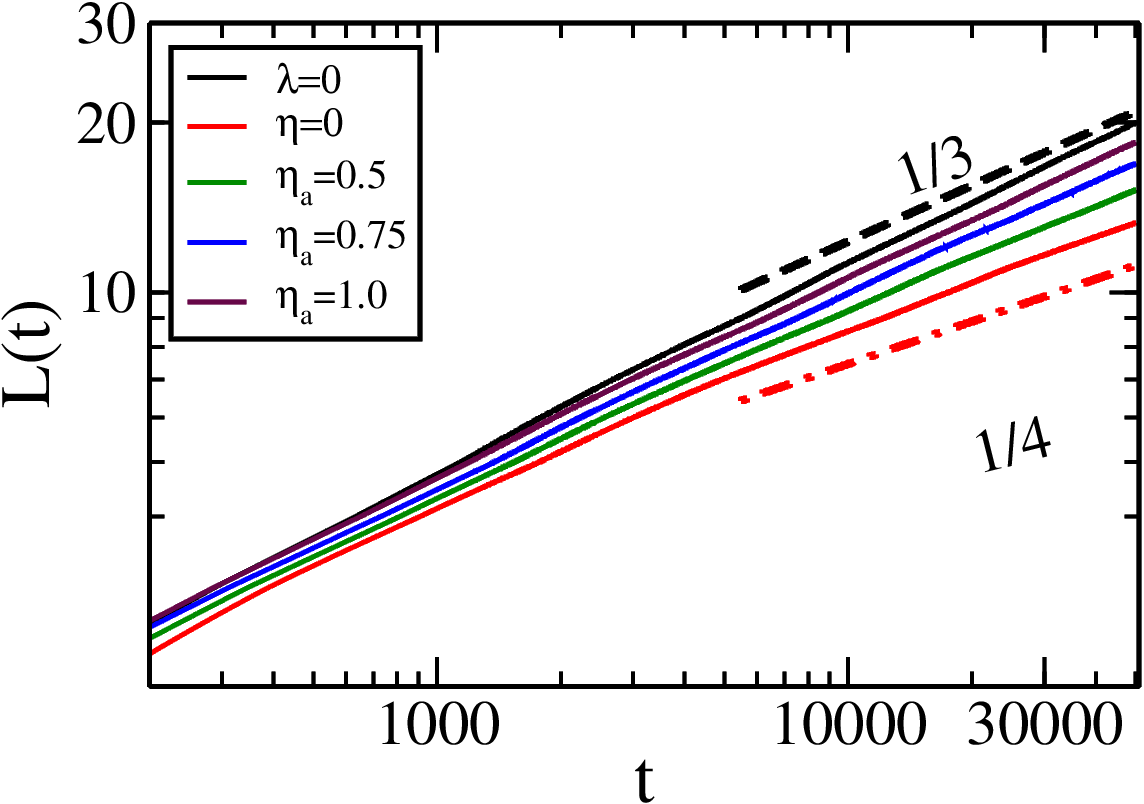}

	\caption{(color online)  Log-Log plot of $L(t)$ vs. $t$ is for different values of $\eta_{a}$. Black and red solid lines refer to the ($\lambda=0$) PMB, and ($\eta=0$) AMB. Lines labeled with $1/3$ and $1/4$ denote the growth laws $L \sim t^{1/3}$ and $L \sim t^{1/4}$, respectively.}
\label{fig:fig2}
\end{figure} 

We numerically integrated the combined eqs.~ \ref{eq:1}, \ref{eq:2} and \ref{eqn_activemodelB}  on a square lattice of size $L \times L$ with periodic boundary conditions in both the directions. We used the Euler discretization scheme with mesh sizes $\Delta x = 1.0$ and $\Delta t = 0.01$. The space mesh is small enough to resolve coarsening interfaces, and the time mesh is adequate to ensure stability of the numerical scheme. The initial condition for a run consist of the order parameter field $\psi ({\bf{r}},0)$ having small-amplitude fluctuations about the mean value $\psi_0 = \int{\psi({\bf r}, t) d {\bf r}} = 0$. This case corresponds to a {\it critical} composition with an equal number of occupied and empty sites. 
Here we only discuss the results of critical composition. The eqs. \ref{eq:1}, \ref{eq:2} and \ref{eqn_activemodelB} are numerically integrated for different values of noise strength $\eta_a \in (0,1)$ and activity parameter $\lambda \in (0,8)$. 
For $\lambda=0$, the model reduces to the passive model B (PMB) \cite{Puri2009,Cahn1959,Hohenberg1977}. For $\eta_{a}=0$ and $\lambda \neq 0$, the model reduces to the previously studied model and referred as  active model B (AMB) \cite{wts14,sudiptapre,sudiptasoft}. The present model is with non-zero noise, $\eta_a \ne 0$  and non-zero activity $\lambda \ne 0$ and we name it active model B with noise (AMBN).

All results presented here are for lattice size $L=512$, and the maximum total time $t=10^5$, where  time $t$ is obtained by multiplying the number of simulation steps with $Delta t$. All the statistical quantities are averaged over $100-150$ independent runs.

\begin{figure}[ht]
\centering
      \includegraphics[scale=0.35]{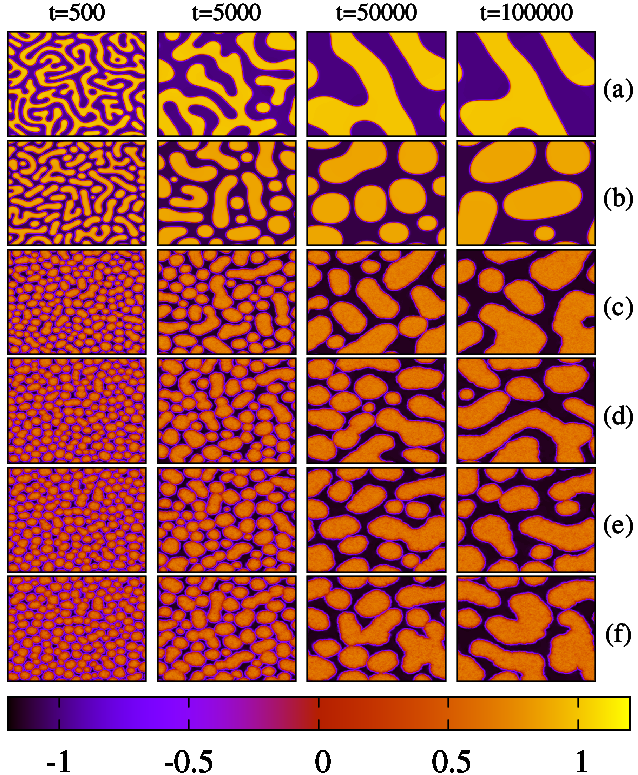}
 
 \caption{(color online) Time evolution of the system is shown for different activities from top to bottom panel. (a-b) indicate activities $\lambda=0$ (PMB),  $\lambda=1$ (AMB), respectively and $\eta_{a}=0$. (c-f)  $\lambda=$ $4$, $6$, $7$ and $8$ respectively, and for a fixed strength of noise $\eta_{a}=0.5$. The color bar shows the range of $\psi$ values. }
\label{fig:fig3}
\end{figure}

\section{Results \label{Results}}
We first start with the results of domain morphology and kinetic of the growing domains 
for the critical composition of the AMBN for different strengths of noise and fixed activity strength $\lambda=1$.
Snapshots of the local order parameter $\psi({\bf{r}},t)$ are shown in Fig. \ref{fig:fig1}, where the color bar represents the $\psi$ value. Blue (dark) and orange (bright) regions refer $\psi({\bf{r}},t)$ is lower and higher 
than the mean density $\psi=0$, respectively. The three columns in Fig. \ref{fig:fig1} are for three different times $t = 5000$, $40000$ and $100000$, and different rows are for different strengths of noise. For all noise strengths, $\eta_{a} = 0$, $0.2$, $0.8$, and $1$ (from rows (a-d) respectively), starting from the initial random homogeneous state, the domains of $\psi$ rich regions start to grow with time. In the absence of noise,  we observe the formation of the isolated domains as shown in Fig. \ref{fig:fig1} (a) (AMB) and also reported in previous studies \cite{sudiptapre, wts14}. 
Furthermore, we note that the system shows the formation of the bi-continuous domains for all noise strengths Fig. \ref{fig:fig1}(c-d). For low noise strength, the bi-continuous nature of domains will appear at later times, whereas for higher noise strength bi-continuous is visible at earlier times, as shown in Fig. \ref{fig:fig1}(d).  

\subsubsection{Domain length and growth exponent \label{Domain length and growth exponent}}

To quantify the kinetic of the growing domains of AMBN, we calculate the order parameter two-point correlation function defined as 
\begin{equation}
 C(r,t)=\langle \psi({\bf r}_{0}+{\bf r},t)\psi({\bf r}_{0},t) \rangle
\end{equation}
                       where the angular bracket $\langle..\rangle$ represents the average over reference positions ${\bf r}_{0}$, and $100-150$ independent realizations. 
                        With time  the system coarsens from the initial random homogeneous state, and the size of the ordered domains increases over time. The domain size is characterized by the characteristic length $L(t)$, which is defined as the distance where the correlation function, $C(r,t)$ drops to $0.1$, of its maximum value at $r=0$.
                      
    We plot $L(t)$ vs. $t$ in fig. \ref{fig:fig2} in log-log scale for $\eta_{a}=0.5$, $0.75$ and $1.0$, keeping the value of activity $\lambda$ fixed to $1.0$. $L(t)$ for PMB and AMB are also included in this plot. It is found that  the ordering kinetics of domain growth for AMB shows a growth law scale as $L(t) \sim t^{1/z}$, where $z$ is the dynamic growth exponent \cite{sudiptasoft, sudiptapre,  wts14, Puri2009}. For PMB $z=3$ \cite{Puri2009}, and for AMB $z$ shows a crossover from $3$ to $4$.  \\
    In our present study with non-zero noise, we find that the asymptotic growth for all noise strength is again similar to as reported for PMB and $z=3$. Although for intermediate times it deviates from $z=3$.

\begin{figure}[ht]
\centering
        \includegraphics[scale=0.21]{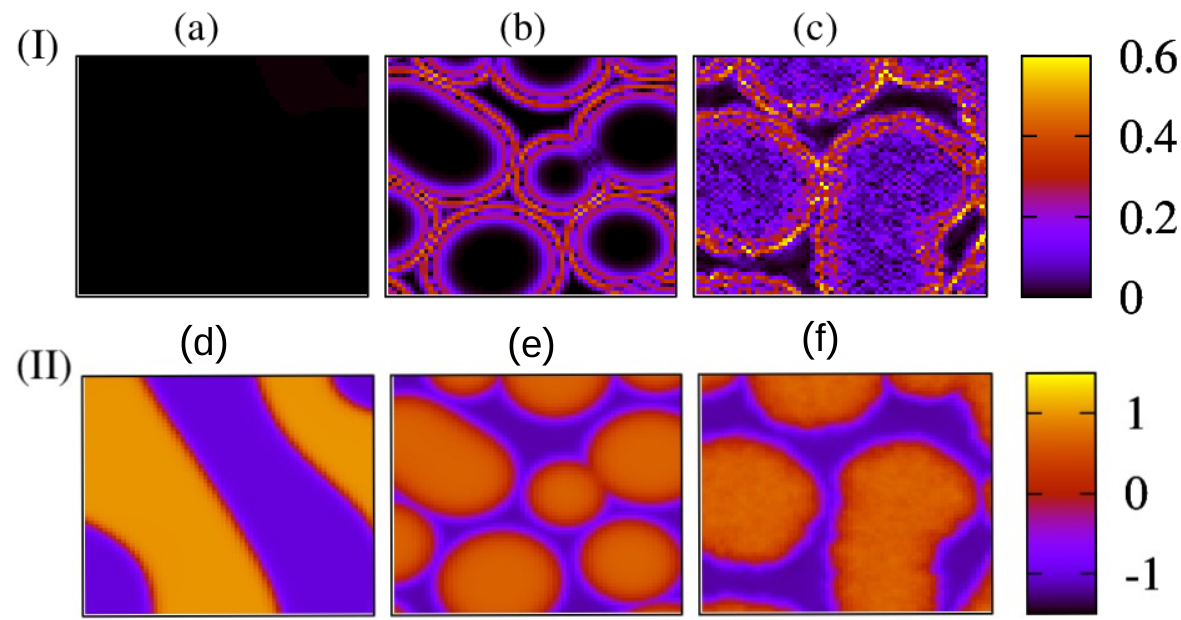}
 \caption{(color online)  Particle current {\bf J}({\bf r}, t) eq. \ref{eq:2} is shown for (I)(a) PMB, (b) AMB, and (c) AMBN. $\lambda$ is fixed at $5$ for AMB and AMBN, and we consider $\eta_{a}=0.5$ for AMBN. (II)(d-f) are snapshots of local $\psi$ corresponding to (a-c) respectively.}
\label{fig:fig4}
\end{figure}            
                       
                       Furthermore, to characterize the time dependence of domain length, we calculate the effective growth exponent, $\frac{1}{z_{eff}}$, as a function of time $t$. Therefore, the $\frac{1}{z_{eff}}$ is  defined as,
\begin{equation}
\frac{1}{z_{eff}}=\frac{d (ln L(t))}{d (ln t)} 
\end{equation}                       

 \begin{figure}[ht]
\centering

 \includegraphics[scale=0.35]{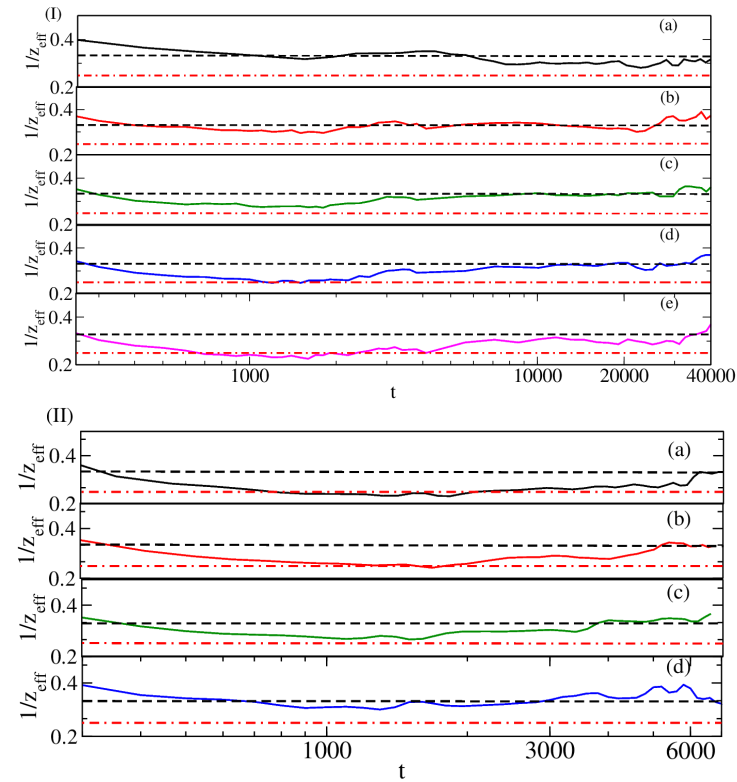}

 \caption{(color online) 
 (a-e) Plots of effective exponent $1/z_{eff}$ vs. $t$ for $\lambda=1$, $4$, $5$, $6$ and $8$, respectively, and $\eta_a=0.5$ for each case. Black and red horizontal dashed lines are drawn at $1/z_{eff}=1/3$ and $1/z_{eff}=1/4$, respectively. (II) (a-d) Variation of effective exponent $1/z_{eff}$ for $\eta_{a}=0.5$, $0.6$, $0.7$ and $0.85$, respectively, and $\lambda=8.0$ is fixed for each case. Black and red horizontal dashed lines are drawn at $1/3$ and $1/4$. }
\label{fig:fig5}
\end{figure}   

          As shown in Fig. \ref{fig:fig2}, the asymptotic growth law is $1/4$ for AMB without noise, and in AMBN the asymptotic growth law is $1/3$. However, when noise and activity are comparable, there is competition between them. This motivated us to explore precisely the behavior of $1/z_{eff}$, and the morphology of domains for intermediate times and high activities. 
             
\begin{figure*}[ht]
\centering
   \includegraphics[scale=0.21]{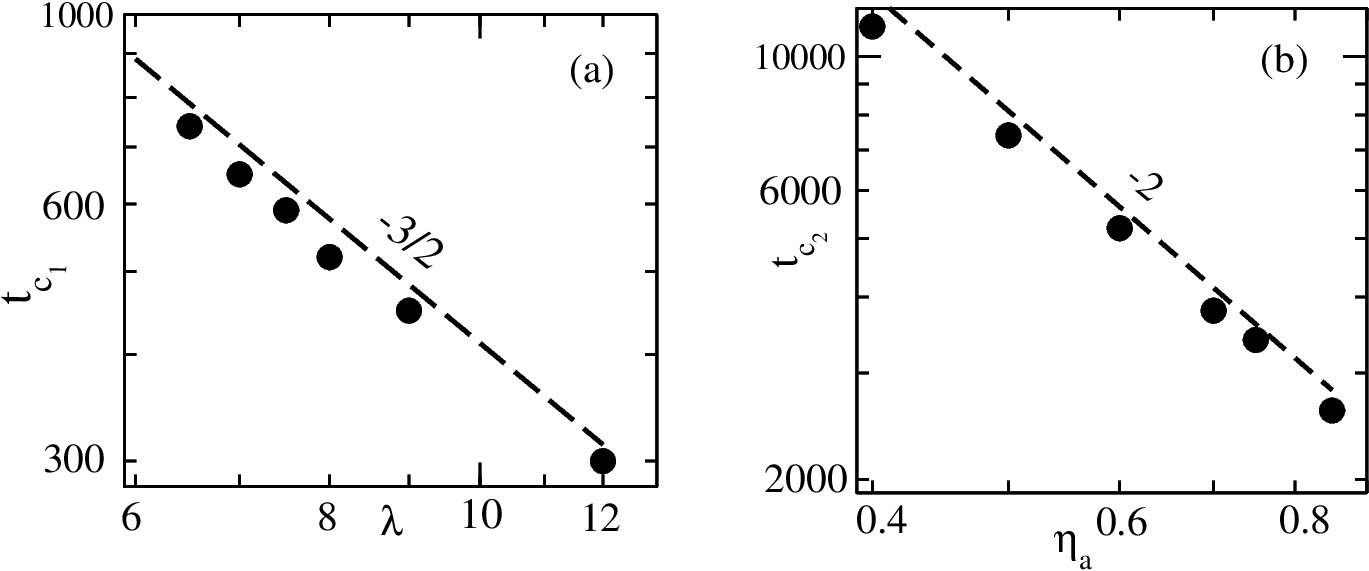} 
   \includegraphics[scale=0.15]{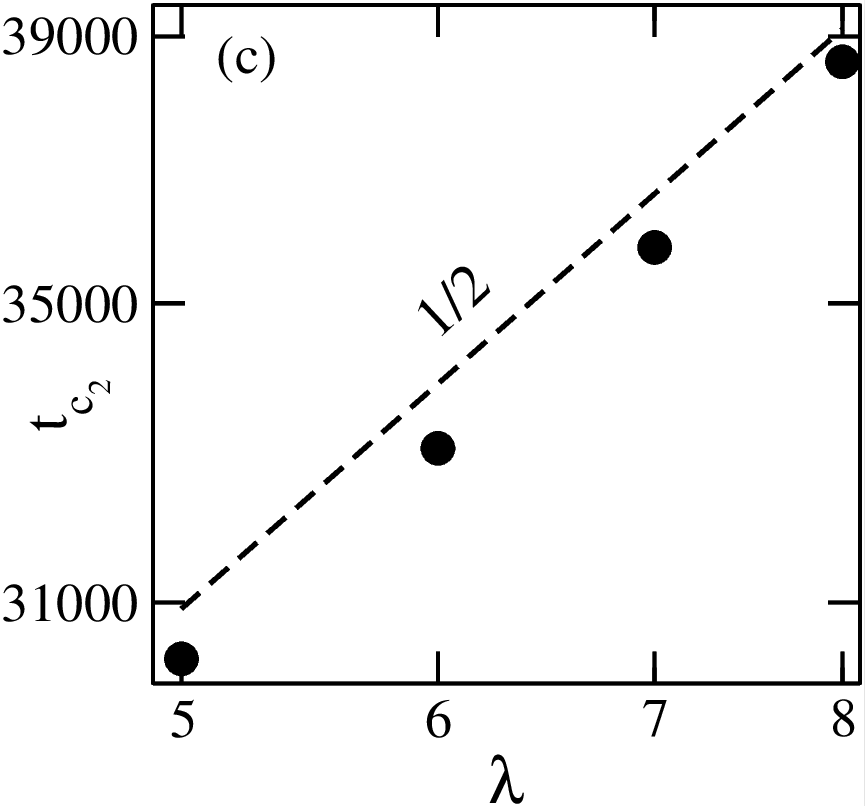} 
   \caption{(color online) (a-c) $Log-log$ plots of $t_{c_{1}}$ vs. $\lambda$, $t_{c_{2}}$ vs. $\eta_{a}$ and $t_{c_{1}}$ vs. $\lambda$ respectively.  Black dashed line refers to slope $-3/2$, $-2$ and $1/2$ in (a), (b) and (c), respectively.}
\label{fig:fig6}
\end{figure*}         
          We further look at the morphology of domains for different activities for a fixed strength $\eta_a=0.5$. In Fig. \ref{fig:fig3}(a-f), we show the real time snapshots of local $\psi$ for four different times $t=500$, $5000$, $50000$ and $100000$, (a-b) for PMB and AMB ($\lambda=1$) without noise, respectively, and (c-f) refer to AMBN with activities $4$, $5$, $6$ and $8$ for fixed noise strength $\eta_a=0.5$.
          For PMB, the structure of the domain is bi-continuous or connected domains for all times Fig. \ref{fig:fig3}(a). For AMB,  Fig. \ref{fig:fig3}(b), the structure of the domains is bi-continuous at early time, as shown for $t=500$ in the figure, whereas it changes to isolated droplet type at late times. 
          In the presence of noise, the early time domains are isolated and they become elongated at late times (last column of (d-f)). We also compare the difference between PMB, AMB, and AMBN by showing the real space snapshots of current ${\bf J}({\bf r}, t)$ (as defined in Eq. \ref{eq:2}), in Fig. \ref{fig:fig4} (I)(a, b, c), respectively, and the corresponding snapshots for local order parameter $\psi({\bf r}, t)$  are shown in Fig. \ref{fig:fig4} (II)(d-f) respectively. For clarity, only a part of the system is shown for each figure. The snapshots of the local order parameter are shown to compare the structure of the interface between high and low-order parameter regions. The color shows the magnitude of current as represented through the color bar.  For PMB,  there is zero current, whereas, for AMB, the current is only present at the interface of the order-disorder region, and the bulk region has almost zero current. The magnitude of current is substantial at the interface. For the AMBN, the disorder region has almost zero current, a large current at the interface, and a small current inside the ordered region of the growing domain. \\
          For comparison, we also performed the microscopic numerical simulation for a collection of active Brownian particles (ABPs) that shows motility induced phase separated (MIPS). We observed that at the boundary of a cluster of ABPs, there is always an exchange of particles with the bulk, and this can result in 
         the presence of non-zero current at the order-disorder interface. Detailed characteristics of such current are the focus of our future study \cite{curr}\\
   
          To further quantify the effect of such current on the kinetics of domain we calculate the $1/z_{eff}$ for the set of parameters $(\lambda, \eta_a)= ([1-8], [0.5])$.  In Fig. \ref{fig:fig5}(I)(a-e), we show the time dependence of $1/z_{eff}$ for different $\lambda=1$, $4$, $5$, $6$ and $8$, respectively, and fixed noise $0.5$. It is found that for $\lambda =6$ and $8$ Fig. \ref{fig:fig5}(I)(d-e),  for early time $1/z_{eff}$ shows a quick approach to $1/4$ and then shows a small plateau at $1/z_{eff}=1/4$ for intermediate times, and then again shows a slow crossover to $1/3$ at late times. As we decrease activity $\lambda$, the $1/z_{eff}$ still shows the early time dip to lower value $1/4$, and it asymptotically approaches to $1/3$, as shown in Fig. \ref{fig:fig5}(I)(b-c). However, for low activity, $\lambda=1$,  $1/z_{eff}$ cannot reach to $1/4$ and system quickly approaches to asymptotic $1/3$ behaviour, as shown in Fig. \ref{fig:fig5}(I)(a). \\

          Similarly we also explore the behaviour of effective growth exponent $1/z_{eff}$ for a fixed activity $\lambda = 8$ and different noise $\eta_a = 0.5$, $0.6$, $0.7$ and $0.85$. In Fig. \ref{fig:fig5}(II)(a-d), we show the comparison plot of $1/z_{eff}$ vs. time $t$. Once again we find competition between activity and noise. For zero noise or for the athermal system, the asymptotic growth is $1/4$, for finite noise and activity the asymptotic growth will become $1/3$ and for the intermediate times, there is competition between the activity and thermal noise. For finite noise and activity and at intermediate times, when domains are small in size and hence the small interface and contribution of the interface current is low and it leads to droplet type domain morphology and $1/4$ growth law. Whereas as the domains become bigger in size and hence the interface, the interface current dominates and lead to the bi-continuous domains as well as the $1/3$ growth at late times.  Hence, depending on noise and activity, the system shows time dependent growth kinetics. 
          Similar domain growth law and crossover from $1/3$ to $1/4$ is previously reported in the context of systems where phase separation occurs via surface diffusion rather than bulk
diffusion \cite{Puri1997, van2005}.  However, surface diffusion cannot be 
the dominant mechanism for the crossover seen in the AMB,
as can be seen by the  droplet morphology of domains as shown in Fig. \ref{fig:fig1}(a). \\
          
\begin{figure*}[ht]
\centering
   \includegraphics[scale=0.355]{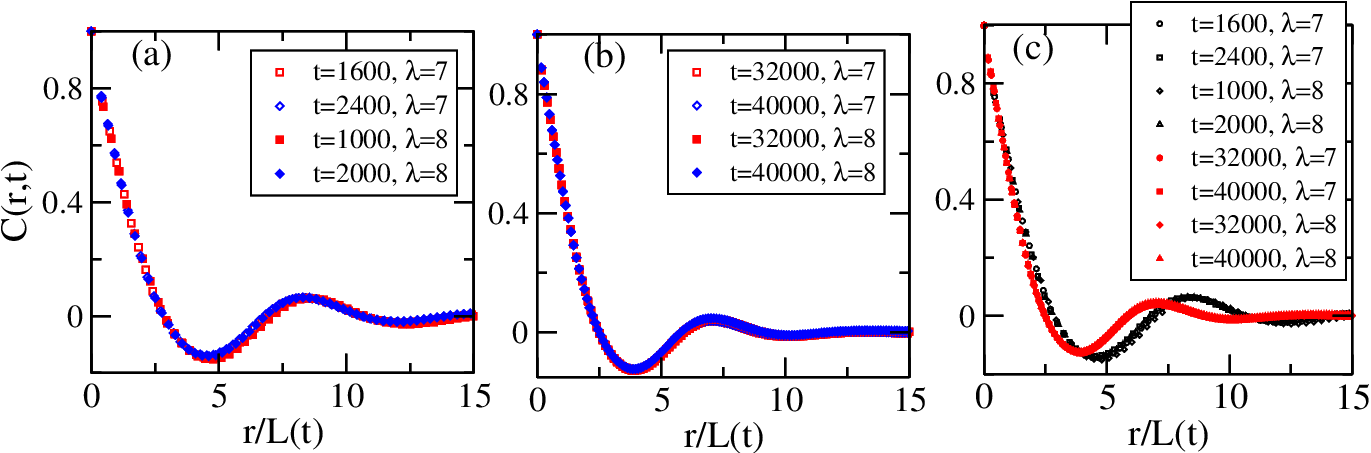}
    \includegraphics[scale=0.355]{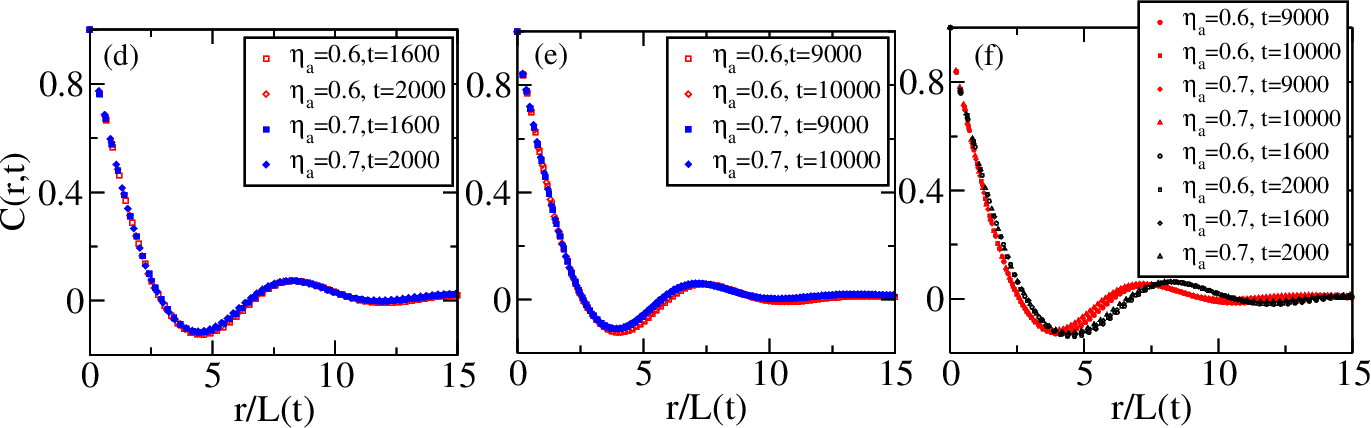}

	\caption{(color online) (a)-(b) Scaled two-point correlation function vs. scaled distance $r/L(t)$ for two different activities $\lambda = 7$ and $8$ for intermediate and late times respectively. (c) is combining all the plots that shows two different scaling.  (d) -(e) are the plots for scaled correlation for fixed activity $\lambda = 8$ and two different noise strengths $\eta_a = 0.6 $ and $0.7$ during intermediate and late times. (f) is combining the plots in same window corresponding to two different scaling.}
\label{fig:fig7}
\end{figure*}

          Further, we calculate the two crossover times $t_{c1}$ and $t_{c2}$. $t_{c1}$ is defined as time when the $1/z_{eff}$, first time crosses from $1/3$ to $1/4$ at early times, this crossover time very much depends on the strength of the activity $\lambda$. During this time the morphology of domains changes from early time bi-continuous type to isolated droplet type as shown in Fig. \ref{fig:fig3}(b-c). The activity dependence of $t_{c1}$ is shown in Fig. \ref{fig:fig6}(a) for fixed noise strength $\eta_a=0.5$. It is clear that $t_{c1} \simeq t^{-3/2}$, the same as obtained in previous study of  AMB \cite{sudiptapre}. Moreover,  in the presence of noise at late times the $1/z_{eff}$ approaches to $1/3$. The second crossover time $t_{c2}$ is defined as the time when $1/z_{eff}$ crosses from $1/4$ to $1/3$ at late times. $t_{c2}$ depends on both the strength of activity and noise. The $t_{c2}$ increases with activity for a fixed noise, and it decreases with noise strength for a fixed activity. In Fig. \ref{fig:fig6}(b-c), we show the dependence of $t_{c2}$ on noise $\eta_a$ and activity $\lambda$. $t_{c2} \simeq \eta_a^{-2}$ and  $t_{c2} \simeq \lambda^{1/2}$.\\
          We can include activity and noise strength dependence of growth law and can write the time dependence of characteristics length with a scaling function for characteristic length which show the limiting behaviour for different parameter regimes, 
          $L(t, \lambda, \eta_a)$
        \begin{equation}
L(t, \lambda, \eta_a) = t^{1/3}F_L(t/t_{c1}, t/t_{c2}) \end{equation}

            the scaling function 
            \begin{equation}
            F_L(x_1, x_2)  \rightarrow a,  x_1 \rightarrow 0, x_2 \rightarrow \infty;
            \end{equation}
            and 
            \begin{equation}
            F_L(x_1, x_2) \rightarrow  b(x_1)^{-1/12}, x_1 \rightarrow \infty, x_2 \rightarrow 0
            \end{equation}
            where, $a$ and $b$ are two constants. \\
            
          Hence, now we clearly see that there are two distinct scaling regimes concerning activity and noise. Now we ask the question; can we have two different scaling functions as well ? For that, we carefully look at the two-point correlation functions for the intermediate and late times.  We know for the appropriate choice of activity and noise, $1/z_{eff}$ remains close to $1/4$ in the intermediate times, and it approaches to $1/3$ at late times or asymptotic time. We plot the scaled two-point order parameter correlation function $C(r)$ as defined before. In Fig. \ref{fig:fig7} (a-b), 
 we show the plot of scaled two-point correlation functions vs. scaled distance $r/L(t)$ for intermediate and late times, respectively, for two different activities $\lambda = 7$ and $8$. We find good dynamic and static scaling in both plots. Furthermore, we combine all the plots of (a) and (b) in Fig. \ref{fig:fig7}(c). For intermediate and late times, we find two distinct scaling functions shown by empty black and filled red symbols, respectively.  Similarly in Fig. \ref{fig:fig7}(d-e), we plot the scaled correlation functions for two different noise strengths $\eta_a = 0.6 $ and $0.7$, respectively, at intermediate and late times, for a fixed activity $\lambda = 8$.   The system shows good dynamic and static scaling for both of them. Then we combine all plots in (d) and (e) and we show two different scaling functions for intermediate and late time by empty black and filled red symbols, respectively in Fig. \ref{fig:fig7}(f). 
          Therefore, we find different scaling regimes for the system depending on activity and noise. The first is when the system crosses from very early time bi-continuous to droplet type, and the growth exponent switches from $3$ to $4$ value. Then, at the asymptotic limit, the domain morphology again becomes bi-continuous, and the growth exponent switches from $4$ to $3$ value.

In the next section we focus our attention to the dynamic scaling theory of  AMBN \cite{Bray1990}.
First we  will show the effect of thermal noise for passive model B and then show that the  thermal noise is relevant for the AMBN.

\subsection{Scaling Theory \label{Scaling Theory}}
{\it Scaling for PMB}:- Following Bray \cite{Bray1990, Brayprl}, we first present the scaling theory for the PMB. We start with the Langevin equation for a conserved density field 
\begin{equation}
 \frac{\partial \psi(r)}{\partial t}=\alpha \nabla^2 \frac{\delta \mathcal{F}}{\delta \psi(r)}+\eta(r,t)
 \label{eqn4}
\end{equation}
For the sake of simplicity, we present the scaling argument for PMB in the case of constant mobility $\alpha$. $\mathcal{F}\{\psi({\bf r}, t)\}$ is the Ginzburg-Landau (GL) free-energy functional \cite{Lubenskey1996}  and $\alpha$ is the mobility factor set to $1$ in eq. \ref{eqn4}.  $\eta(r,t)$ is a Gaussian white noise generated by thermal fluctuations with mean zero and correlations
$ <\eta(r,t)\eta(r',t') >  = 2 \alpha T \nabla^2\delta(r-r')\delta(t-t')$. 
Introducing the Fourier transform
$\psi_{k}(t)= V^{-1/2} \int d^{d}r  \psi(r,t)\exp(i{\bf k}.{\bf r})$, $V$ is the area of  space, the equation in Fourier space we get 
\begin{equation}
\frac{\partial \psi_{k}}{\partial t}=-k^2 \alpha \frac{\delta \mathcal{F}}{\delta \psi_{-k}}+ \eta_{k}
\end{equation}
\begin{equation}
\frac{1}{\alpha k^2} \frac{\partial \psi_{k}}{\partial t}=-\frac{\delta \mathcal{F}}{\delta \psi_{-k}}+ \xi_{k}(t)
\end{equation}
                      where $\xi_{k}(t)=\frac{\eta_{k}}{\alpha k^2}$.
Introducing rescaling of space and time,
$k=k'/b$ and $t=b^zt'$ leads $\psi_{k}(t)=\psi_{k'/b}(b^zt')= b^\zeta \psi'_{k'}(t')$ and
$\mathcal{F}(\psi_{k})=b^y\mathcal{F}(\psi'_{k'})$

which gives
\begin{equation}
\frac{b^{2-z+\zeta}}{\alpha k'^2} \frac{\partial \psi'_{k'}}{\partial t'}= -b^{y-\zeta} \frac{\delta \mathcal{F}'}{\delta \psi'_{-k'}}+ \xi_{k'/b} (b^zt')
\end{equation}

\begin{equation}
\frac{b^{2-z-y+2\zeta}}{\alpha k'^2} \frac{\partial \psi'_{k'}}{\partial t'}=  -\frac{\delta \mathcal{F}'}{\delta \psi'_{-k'}}+b^{-y+\zeta} \xi_{k'/b} (b^zt')
\label{eq:eq13}
\end{equation}

\begin{equation}
\frac{b^{2-z-y+2\zeta}}{\alpha k'^2} \frac{\partial \psi'_{k'}}{\partial t'}=  -\frac{\delta \mathcal{F}'}{\delta \psi'_{-k'}}+ \xi'_{k'} (b^zt')
\end{equation}
where the new noise $\xi'_{k}(t)$ is
\begin{equation}
<\xi'_{k'}(t_{1}) \xi'_{-k'}(t_{2})>= b^{2-z+2\zeta-2y} \frac{2T}{\alpha k'^2} \delta (t'_{1}-t'_{2})
\label{eq:eq15}
\end{equation}
 
\begin{equation}
\frac{1}{\alpha' k'^2} \frac{\partial \psi'_{k'}}{\partial t'}= -\frac{\delta \mathcal{F}'}{\delta \psi_{-k'}}+\xi'_{k'}(t')
\end{equation}

where $<\xi'_{k'}(t'_{1}) \xi'_{-k'}(t'_{2})>=  \frac{2T'}{\alpha k'^2} \delta (t'_{1}-t'_{2})$ and $(\frac{1}{\alpha})'=(\frac{1}{\alpha})b^{2-z-y+2\zeta}$ (from eq \ref{eq:eq13})

Putting the value of $1/\alpha$ in equation \ref{eq:eq15} we have

\begin{equation}
<\xi'_{k'}(t_{1}) \xi'_{-k'}(t_{2})>= \frac{2Tb^{2-z+2\zeta-2y}}{\alpha'b^{2-z-y+2\zeta}k'^2} \delta(t_{1}-t_{2})
\end{equation}

this yields
\begin{equation}
<\xi'_{k'}(t_{1}) \xi'_{-k'}(t_{2})>=\frac{2Tb^{-y}}{\alpha'k'^2} \delta(t_{1}-t_{2})
\end{equation}

So, from here we obtain that
\begin{equation}
T'=Tb^{-y}
\end{equation}
where, $b>1$ and $y>0$, so $T'\rightarrow 0$  and we find temperature is irrelevant \cite{Bray1990}.   \\\par
 {\it Scaling for  AMBN}:-
We start with the continuity equation for the local order parameter $\psi$  as given in eq. \ref{eq:1}, 
\begin{equation}
\begin{aligned}
\begin{split}
\frac{\partial}{\partial t} \psi({\bf{r}},t) &= -{\bf{\nabla}} \cdot \left[{\bf{J}}({\bf{r}},t)  + {\bf n}({\bf{r}},t)\right] ,   \\
\label{eq:11}
\end{split}
\end{aligned}
\end{equation}
 where $\bf{J}({\bf{r}},t)$ is as given in eq. \ref{eq:2}. For simplicity we perform the scaling theory for constant mobility in eq. \ref{eq:2}. Because as shown previosuly \cite{Puri1997, van2005}, the density dependence of the mobility is important where  phase separation occurs via surface diffusion rather than bulk diffusion. But surface diffusion is clearly not the important mechanism in AMB \cite{sudiptapre}.  But due to presence of finite current across the interface in AMB system as shown in Fig \ref{fig:fig4}(I)(b) , noise ${\bf n}$ should me multiplicative in nature. \\
Further splitting in chemical potential $\mu({\bf r}, t)$ in two parts, we can write the equation \ref{eq:11} as
 \begin{equation}
 \frac{\partial \psi}{\partial t}= \alpha \nabla^2 \frac{\delta \mathcal{F}}{\delta \psi}+ \lambda \nabla^2 (\nabla \psi)^2 + \nabla.{\bf n}({\bf{r}},t)
 \label{eq:12}
 \end{equation}
 The multiplicative noise {\bf n}({\bf{r}},t) is the same as defined in eqs. \ref{eq:4} and \ref{eq:noise}. The multiplicative noise in AMBN as given in eq. \ref{eq:noise}  has mean zero and strength proportional to $ \eta \backsimeq$ $\eta_{a}\psi^{3/2}$. Further we can write the eq. \ref{eq:12} as
 \begin{equation}
 \frac{\partial \psi}{\partial t}= \alpha \nabla^2 \frac{\delta \mathcal{F}}{\delta \psi}+ \lambda \nabla^2 (\Lambda(\nabla \psi)) + \nabla.{\bf n}({\bf{r}},t)
 \label{eq:13}
 \end{equation}
 where $\Lambda(\nabla \psi)$ is the term due to activity and is a functional of gradient of $\psi({\bf r}, t)$.
Making the transformation from real space to Fourier space, the Fourier transformed equation will be, 
 \begin{equation}
\frac{\partial \psi_{k}}{\partial t}= -\frac{\alpha k^2 \delta \mathcal{F}}{\delta \psi_{-k}}+ \lambda k^2 \Lambda_{k} +\eta_{k}
\end{equation}
We can further divide the whole equation by $\alpha k^2$, such that the coefficient is $1$ in front of $\frac{\delta \mathcal{F}}{\delta \psi_{-k}}$. 
\begin{equation}
\frac{1}{\alpha k^2} \frac{\partial \psi_{k}}{\partial t}= -\frac{\delta \mathcal{F}}{\delta \psi_{-k}}+ \frac{\lambda}{\alpha}  \Lambda_{k}(t) +\xi_{k}(t)
\end{equation}
where the new modified noise $\xi_{k}(t)=\frac{\eta_{k}}{\alpha k^2}$.
We start with rescaling of space and time as before $ k=\frac{k'}{b}$, $t=b^zt'$, $\psi_{k}(t)=b^\zeta \psi_{k'}(t')$, $\mathcal{F}(\psi_{k})=b^y\mathcal{F}(\psi'_{k'})$ and assuming that $\Lambda_k = \Lambda_k'$.

\begin{equation}
    \frac{b^{2-z-y+2\zeta}}{\alpha k'^2} \frac{\partial \psi_{k'}}{\partial t'} = - \frac{\delta \mathcal{F}'}{\delta \psi'_{-k'}}    \\
+\frac{\lambda}{\alpha}b^{\zeta-y}\Lambda_k' + \xi'
\label{eq:23}
\end{equation} 

\begin{equation}
    \frac{1}{\alpha'} \frac{\partial \psi_{k'}}{\partial t'} = - \frac{\delta \mathcal{F}'}{\delta \psi'_{-k'}}    \\
+\frac{\lambda}{\alpha}b^{\zeta-y}\Lambda'_k+\xi'
\label{eq:24}
\end{equation} 

   where 
\begin{equation}
 <\xi'(t'_1) \xi'(t'_2)> = b^{-2y+2\zeta}\frac{2Tb^2}{\alpha k'^2}b^{3\zeta-z}\delta(t'_{1}-t'_{2})  
\end{equation}   
  
  this gives 
  \begin{equation}
  <\xi'(t'_1) \xi'(t'_2)> = b^{-2y+5\zeta+2-z}\frac{2T}{\alpha k'^2}\delta(t'_{1}-t'_{2})      
  \end{equation}

From eq \ref{eq:24} , we can obtain;
$\frac{1}{\alpha'}=\frac{1}{\alpha}b^{2-z-y+2\zeta}$
hence
\begin{equation}
<\xi(t'_1) \xi'(t'_2)>=\frac{b^{5\zeta-2y+2-z}}{\alpha'b^{2-z-y+2\zeta}}\frac{2T}{k'^2}\delta(t_{1}'-t_{2}')
\end{equation}
this gives

\begin{equation}
T'=Tb^{3\zeta-y}
\end{equation}
Imposing the condition  that  the temperature is irrelevant lead to  $3\zeta-y <0$, which implies $3\zeta<y$ and if we take $\zeta=\frac{d}{2}$ and $y=d-1$ from equilibrium theory \cite{Cahn1959} we get 
$\frac{3d}{2} < d-1$, so   we can argue that the temperature is no longer an irrelevant variable for AMBN. 


 
\section{Discussion \label{Discussion}}
We perform  the numerical study and scaling theory to understand the  kinetics of active model B with thermal noise. The thermal noise in active model B, is multiplicative as a result of local density dependence of motility of particles.  Such noise becomes relevant due to the presence of finite current at the order-disorder interface, whereas the current is zero for corresponding passive model B. The existence of non-zero current can also be observed due to the dynamic nature of phase-separated clusters in ABPs system.   \\
In the presence of thermal noise, the size of the ordered domain in AMB shows a strong time-dependent growth law. When quenched from a random disordered state to an ordered state for AMB,  
the size of the ordered domains increases with time with an effective asymptotic growth exponent $4$ as found in our previous study \cite{sudiptapre}. Unlike the PMB, we find that noise or temperature is relevant for AMBN. Using extensive numerical study, we find that the noise changes the asymptotic effective growth exponent from $4$ to $3$ in AMBN, and the structure of the domains changes from spherical isolated domains to bi-continuous elongated domains. However, for a fixed noise and activity, the size of the domain grows with a time-dependent growth exponent. For early to intermediate times, the growth exponent shows a crossover from early times $3$ value to intermediate times $4$, and finally, at late time it again approaches to $3$. Correspondingly we note  two different scaling functions for intermediate time with growth exponent $4$ and late time with growth exponent $3$.

Furthermore, the role of noise in the system away from the critical composition is also an interesting problem to study in future works. Since the noise is intrinsic to real experimental systems. The role of noise in ordering kinetics in active models is an interesting direction to understand the effect of thermal fluctuations in other active systems \cite{Cates2018, Maiti2019, Chattopadhyaypre, Chattopadhyay2023, Toner2012}. 

\section{Acknowledgement}
S.D. acknowledges the support and the resources provided by PARAM Shivay Facility under the National Super computing Mission, Government of India at the Indian Institute of Technology, Varanasi. I.I.T. (BHU) Varanasi. S.M. thanks DST, SERB (INDIA), MTR/2021/000438, and CRG/2021/006945 for ﬁnancial support.

\end{document}